\documentclass[12pt,preprint]{aastex}

\shorttitle{Element enhancements along the entire AGB phase}
\shortauthors{Wylie-de Boer and Cottrell}

\begin{document}


\title{Element enhancements along the entire AGB phase}

\author{E. C. Wylie-de Boer}
\affil{Research School of Astronomy and Astrophysics, Australian National University, \\ Mount Stromlo Observatory, Cotter Rd, Weston Creek, ACT 2611, Australia\\and\\Department of Physics and Astronomy, University of Canterbury,\\ P.O. Box 4800, Christchurch, New Zealand}
    \email{ewylie@mso.anu.edu.au}
\and
\author{P. L. Cottrell\altaffilmark{}}
\affil{Department of Physics and Astronomy, University of Canterbury,\\ P.O. Box 4800, Christchurch, New Zealand}
\email{peter.cottrell@canterbury.ac.nz}


\begin{abstract}

The results of a study of the AGB phase of stellar evolution are presented.  Abundances have been determined for Fe, C, O, the light s-process elements, Y and Zr, the heavy s-process elements, La and Nd, and the r-process element, Eu.  The expected relationship between enhanced C, increasing C/O ratio and enhanced s-process elements has been quantified.  Results are presented to provide observational data with which to compare theoretical predictions.  

The results in this paper confirm previously suggested relationships between C, C/O and s-process element enhancements.  It is seen that AGB stars show C/O ratios from C/O$\sim$0.4 to 1.0, while C enhancements lie between [C/Fe]=0.1 to 0.9 dex.  Enhancements of s-process elements are as much as [s/Fe]$\sim$1.0 dex for the stars in which C is also greatly enhanced.

 \end{abstract}


 
\section{Introduction}

The analysis of the surface abundances of asymptotic giant branch (AGB) stars offer tests for the operation of the third dredge-up (TDU).

The surface of AGB stars can be enriched in carbon as a direct result of the TDU phenomenon (Lattanzio and Karakas, 2001).  Stars are formed with a C/O ratio $<$1 but at the end of the AGB the stars can show a C/O ratio $>$1 in the envelope as a result of this mixing process that pollutes the envelope with carbon-rich material.  The stellar mass plays a critical role in determining the enhancement of both carbon and the s-process elements, as it directly affects both the source of neutrons in the stellar interior and the efficiency of the TDU episodes (see Busso et al., 1999, and references therein).  The final C/O ratio is also affected by the mass loss history although a discussion of this is not given here.  Thorough details of observations of AGB stars are vital to enable comparison with the enrichments predicted by numerical stellar nucleosynthetic modeling.  Stars can either enrich themselves via their own internal TDU, in which case they are known as \textit{intrinsic} carbon stars, and display technetium in their spectra, or are enriched from transfer of carbon-rich material from a binary companion, in which case they are known as \textit{extrinsic} carbon stars and any (radioactive) technetium will have decayed into a stable isotope of ruthenium.  

The carbon content of the stellar envelope is expected to increase
with changing spectral type along the AGB, from M to MS to S to C, where M stars show C/O$\sim$0.5 while C stars can show C/O$>$1.  This TDU enrichment process also acts to increase the envelope abundances of s-process elements.  Light s-process (Y and Zr) and heavy s-process (La and Nd) elements discussed in this study are hereafter referred to as `ls' and `hs' respectively. 

The spectra of M stars are classified by strong absorption bands of TiO and by the large number of strong metallic lines blueward of 4000\AA (Jaschek and Jaschek, 1987).  The strength of the TiO bands increases rapidly with advancing subtype, and continuum fitting becomes difficult.  Near the end of the M classification, bands of VO become strong compared to TiO.  MS stars are placed between the separate M and S classifications, and display molecular bands of both ZrO and TiO, although the latter still dominates.  

S stars have a C/O ratio approaching unity.  They are late-type stars showing distinct bands of ZrO in the blue and visual.  The easiest way to distinguish S stars is the comparison between strengths of the ZrO and TiO bands, with ZrO taking over as the S subclass increases.  

SC stars are thought to lie between S and C type stars and show characteristics of both classes.  There is some observational evidence that SC stars can be formed in two different ways; either through the continuous dredge-up of C to the envelope during the AGB phase, or due to the hot bottom burning phase in massive AGB stars.  In either case, the phase during which C/O $\sim$ 1.0 is very short, which may explain why the number of confirmed SC stars is so low (Guandalini and Busso, 2008).  They are classified by their C/O ratio of $\sim$1, with most of the C and O locked up in CO.  They exhibit enhanced heavy metal lines, and bands of CN are prominent.  They also display very strong
NaD lines and weak but clear C$_2$ and ZrO bands.  Due to the balance of oxygen and
carbon, all oxide bands and carbon molecules are weak and the
spectra are full of atomic lines, making them easier targets for
abundance analysis.  However, only a few in-depth studies on heavy
element abundances have been undertaken (Catchpole, 1982; Abia and Wallerstein, 1998).  

C stars show very crowded spectra due to strong molecular bands and low
temperatures.  Their C/O ratios are very similar to those shown by SC
stars.  Again, only a few studies on heavy element abundances have
been undertaken (Utsumi, 1970; Abia et al., 2001; Abia et al, 2002).

In order to compare observed s-process enhancements with those predicted by theory, it is desirable to have a set of results along the AGB phase that are internally self-consistent.  Relatively few studies have been devoted to large samples of AGB stars, usually studying only one or two spectral classes at a time. 

A sample of 22 M and MS stars have been analysed by Smith and Lambert (1985, 1986).  These studies used high-resolution, infra-red spectra to obtain abundances for C, N, O, ls and hs elements.  They found s-process enhancements in seven of the eleven M stars but no measurable enhancements in four of the eight MS stars.

The most detailed study of S and SC stars has been undertaken by Abia and Wallerstein (1998) using high-resolution spectra over the optical wavelength range.  This study of S and SC stars derived abundances for C, N, O and a large number of heavy elements.  They found considerable s-process elements enhancements in all S and SC stars, of values as much as [s/Fe]$\sim$+1.4 dex.

Both of these studies measured equivalent widths for atomic lines, including s-process lines, which are often weak and are affected heavily by molecular blending.

This present study of eleven stars has eight stars in common with these previous studies, and includes three stars not previously analysed.  It provides a self-consistent set of results over a range of spectral classes, M, MS, S and C using high-resolution (R$\sim$40,000), high signal to noise (S:N$>$100) spectra over an optical wavelength range of 4500 to 7500\AA.  The self-consistency of these results minimises all discrepancies due to differing methods or systematic offsets between studies.  It allows for an internally accurate abundance analysis of stars relative to each other along the AGB phase.  Observations and analysis of this sort are of great importance to compare to the enhancements predicted by stellar models in order to refine various variable model parameters.  Full details of the method of analysis are discussed in Section 2. The results of the abundance analysis are presented in Section 3, before a discussion of both specific results and general trends in Section 4.  A brief conclusion is given in Section 5. 

\section{Observations and Analysis}\label{obsan}

The eleven field AGB stars analysed are listed in Table \ref{starquant}, with previously published spectral types (Smith and Lambert, 1986), recent magnitudes (SIMBAD database\footnote{http://simbad.u-strasbg.fr} and 2MASS survey\footnote{http://irsa.ipac.caltech.edu}), and calculated bolometric magnitudes, variability type (Samus and Durlevich, 2004) and previous detection of the radioactive element, technetium (Smith and Lambert, 1988).  Bolometric magnitudes were calculated using the bolometric correction approximation from Alonso et al.(1999) and absolute K magnitudes using SIMBAD's apparent magnitudes and distances from the HIPPARCOS catalogue \citep{hip}.  Magnitudes are not given for HD286340 due to its extreme variability preventing reliable measures of its brightness. 

Observations were taken with the 1m telescope and HERCULES spectrograph at Mt John University Observatory, New Zealand, over 2004 September, and the Anglo-Australian Telescope and UCLES at Siding Springs over 2004 August.  Standard reduction procedures in FIGARO were followed for all observations. 

The abundance analysis for these field stars included using atomic line lists from Kurucz\footnote{http://kurucz.harvard.edu} with refined oscillator strength values for the s-process lines via a reverse solar analysis.  This reverse solar analysis was done with a solar model of T$_{eff}$=5770K and log g=4.44.  A similar reverse analysis was also undertaken using Arcturus as this red giant more closely represents the atmospheres dealt with in this research and lines too weak for detection in the sun are often seen at these cooler temperatures.  The Arcturus model adopted was T$_{eff}$=4300K, log g=1.50 and [Fe/H]=-0.50.  Due to the fact that at these cooler temperatures and in these C enhanced stars, molecular features become very strong, all molecular features of CO, CN, CH, ZrO, VO, and TiO were included.  With the exception of TiO line data, which was taken from Plez (1998), all molecular line data was taken from Kurucz's linelists.  This study does not employ equivalent width measurement for any lines other than iron due to strong molecular blending and difficult continuum placement.  All abundances are obtained using the spectrum synthesis program MOOG (Sneden, 1973).  The Fe lines used for equivalent width analysis to determine the atmospheric parameters (see Section 2.1) are listed in Table \ref{felinelist} while the oxygen and heavy element lines used for spectrum synthesis analysis are listed in Table \ref{otherlinelist}.  This spectrum synthesis program adopts solar abundances from Anders and Grevesse (1989).  Of specific note are the carbon, nitrogen and oxygen abundances, which were taken as log($\epsilon$)C=8.56, log($\epsilon$)N=8.05 and log($\epsilon$)O=8.93.  
 
\subsection{Atmospheric Parameter Determination}
Due to severe molecular blending and difficult continuum placement in these stars, it becomes increasingly difficult to find unblended Fe I and Fe II lines for equivalent width measurement with which to determine the atmospheric parameters.   However, this technique was adopted for all stars in an attempt to obtain a spectroscopic effective temperature and gravity.  With a total of around 28 Fe I lines and 8 Fe II lines, the usual comparison between excitation potential and derived abundance was used to determine the effective temperature while the balance between abundances derived from neutral and ionised iron lines was used to refine the gravity of the star.  The microturbulent velocity was obtained by measuring Fe I lines and ensuring that the derived abundances be independent of equivalent width.  

The final spectroscopic effective temperatures ranged from 3650K to 4000K.  A photometric temperature was also calculated using the colour-index calibration published by Alonso et al. (1999), assuming an [Fe/H] of 0.0 for this calibration (see Table \ref{fsap}), which ranged from 3640K to 3970K.  With three exceptions, HD7351, HD49331 and HD64332, the spectroscopic and photometric temperatures agree to within 50K.  All gravities fell between the expected values of log g=0.5 to 1.5.  For one star, HD286340, no reliable B-V values could be found due to the star's extreme variability.  In this case, typical values of T$_{eff}$=3500K, log g=1.0 and $\xi$=1.50 km s$^{-1}$were chosen.  All atmospheric parameters derived are shown in Table \ref{fsap}.  Model atmospheres are taken from the MARCS model atmosphere grid (Gustafsson et al, 2008).  C-rich MARCS model atmospheres were adopted for those stars in which C/O$\simeq$1.0 (HD 30959, HD 64332 and HD 286340).

\subsection{C and O Line Detail}
In addition to the s-process element abundances, the C and O abundances are of relevance along the AGB phase.  Given that the C/O ratio changes along the AGB, due to the dredging up of C after the thermal pulses, it is expected that a correlation should exist between [C/Fe], C/O and [s/Fe].  In order to quantify this relationship, C and O abundances were determined for these field stars.  

The C abundance was determined from the C$_2$ bands in the 5060 to 5100\AA\ region.  The nature of C$_2$ results in extreme sensitivity to any change in C abundance, making it a reliable molecule to use in deriving any C enhancements.  The C$_2$ molecular line parameters were taken from Kurucz's linelists.

The O abundance was obtained from spectrum synthesis of the forbidden O lines at 6300 and 6363\AA.  The oscillator strength of the forbidden line at 6300\AA\ is well known, and a value of log(g)=-9.750 was adopted (as discussed in Lambert, 1978).  The 6363\AA\ O line is very weak and should not be used for abundance determination on its own.  However, in this study it was used as a check of the abundance derived from the 6300\AA\ line wherever it was detectable.  The oscillator strength adopted was log(g)=-10.25 (Lambert, 1978).  These values differ from those supplied in Kurucz's line list by only -0.07 and -0.05 respectively.

\section{Results}
\label{results}
All abundance results for this sample of field AGB stars are shown in Tables \ref{mstars} and \ref{sstars}.  Specific results of interest are discussed further below.

\subsection{Fe abundance}
Iron abundances were obtained for all stars for which spectroscopic atmospheric parameters could be determined.  The solar iron abundance adopted for all abundance analyses was log($\epsilon$)Fe = 7.52, close to the meteoritic abundance published by Anders and Grevesse (1989).  For HD286340, for which standard parameters were adopted, an iron abundance of [Fe/H]=0.0 was assumed.  [Fe/H] values ranged from -0.26 dex to +0.17 dex.  Thus all stars analysed can be considered of solar or near-solar metallicity.  All stars are members of the disk population, and the observed spread in [Fe/H] in the sample results from both uncertainties and genuine differences.  Of the eight stars in common with previous studies (Smith and Lambert, 1985; Smith and Lambert, 1986; Abia and Wallerstein, 1998), all but one [Fe/H] value agrees with previous reported values  to within 0.10 dex. 

\begin{figure}
\plotone{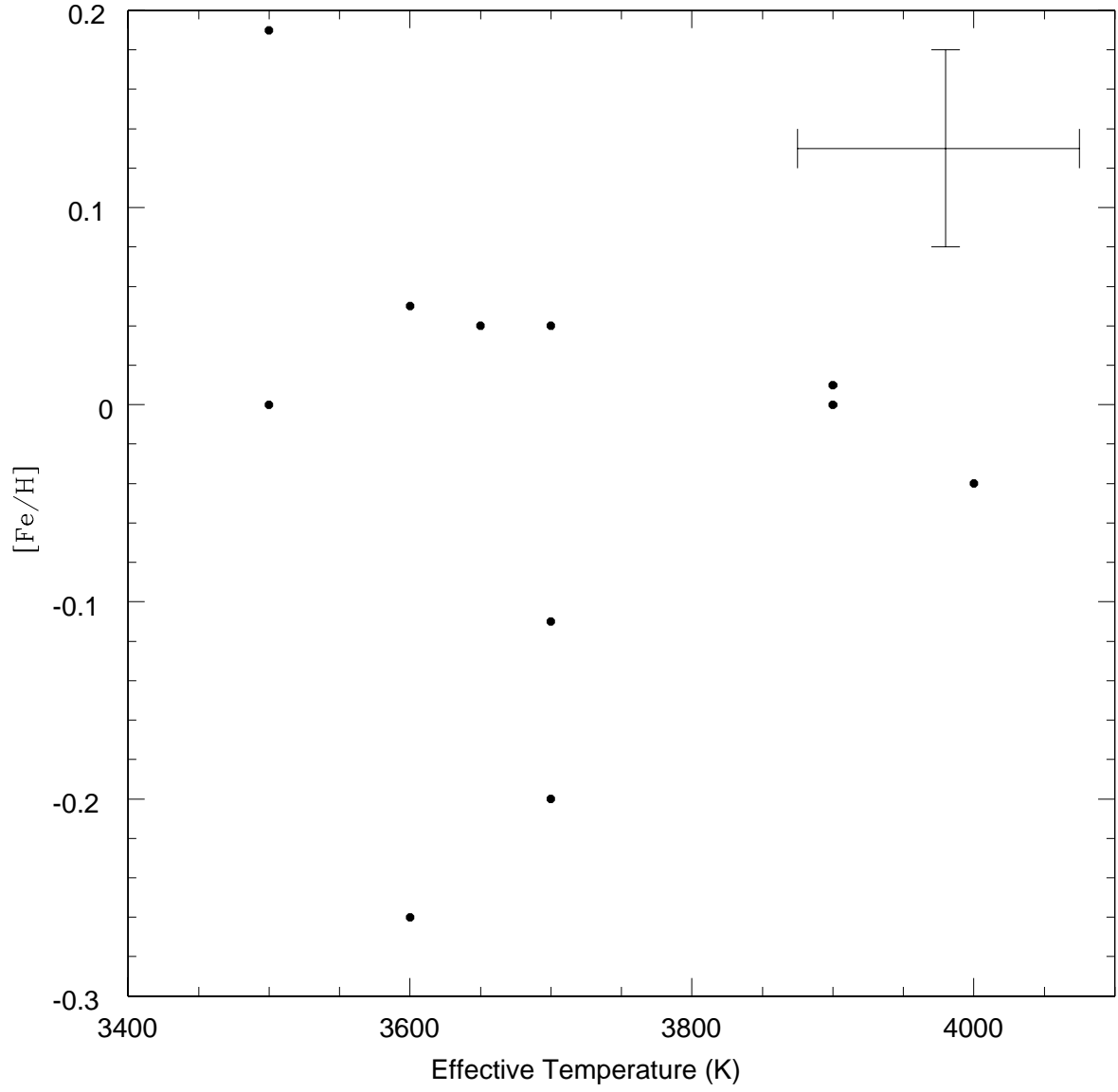}
\caption{Relationship between [Fe/H] and effective temperature for field stars.} \label{feteff}
\end{figure}

As a check of the non-LTE effects on low-excitation potential neutral lines the iron abundance was plotted against effective temperature, as seen in Figure \ref{feteff}.  No real trend is seen for [Fe/H] as a function of temperature.  The Fe I lines used for effective temperature determination have medium to high excitation potentials, $\chi=2.5-5.0$eV, and are thought to be less affected by dramatic non-LTE effects.

\subsection{C and O abundances}
\begin{figure}
\plotone{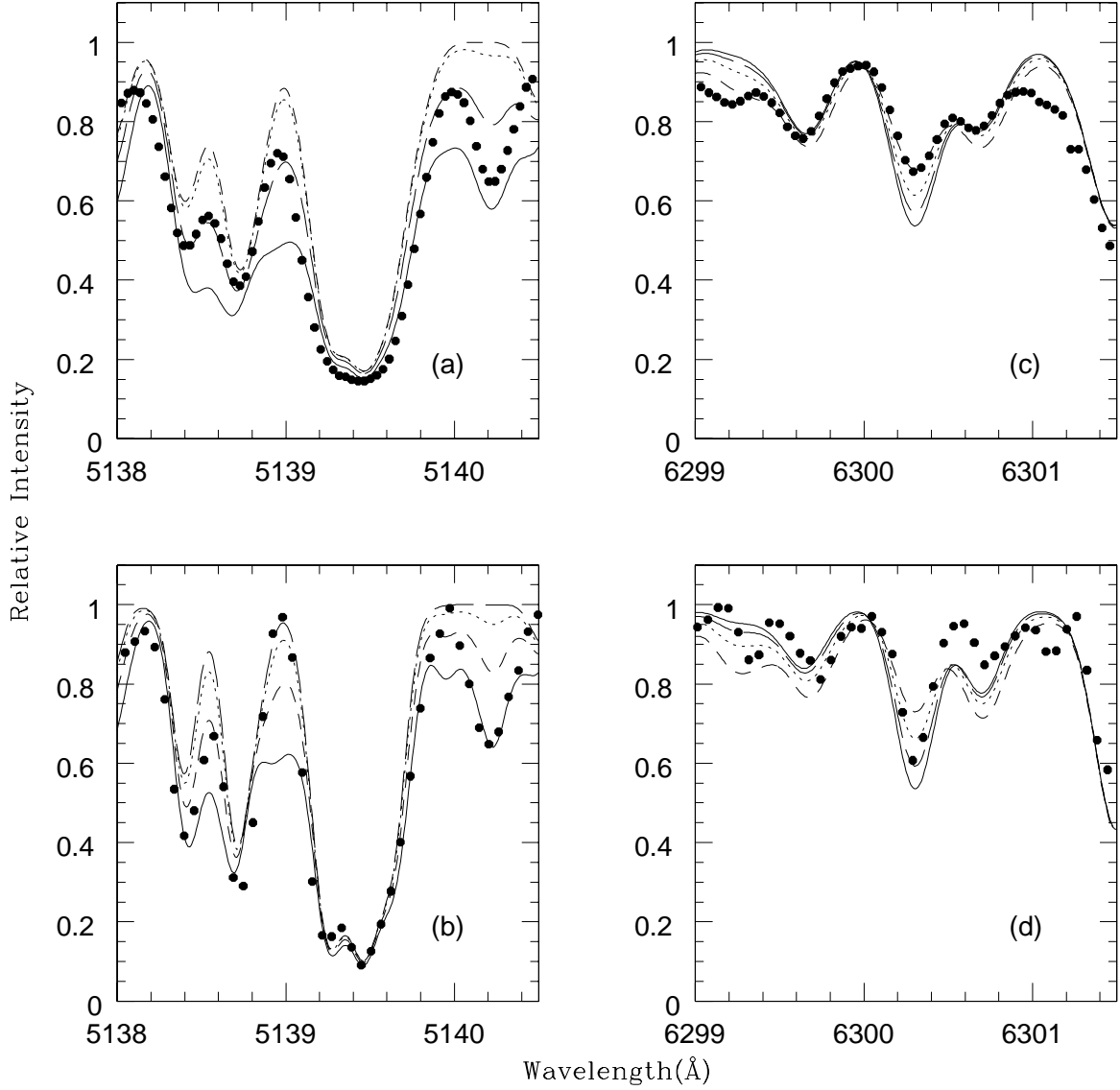}
\caption{Examples of line synthesis used to ascertain C and O abundances.  a) HD49331 with C abundances of [C/Fe]=0.00 (short dashed), 0.05 (dotted), 0.15 (long dashed) and 0.30 (solid) and b) HD64332 with C abundances of [C/Fe]=0.00, 0.50, 0.65 and 0.80; c) HD49331 and d) HD64332 with O abundances of  [O/Fe]=0.00, 0.10, 0.25 and 0.40.} \label{syncofield}
\end{figure}

As mentioned earlier, the abundances of C and O in AGB stars are of interest when comparing with theoretical predictions.  It is expected that the C abundance will increase with recurring thermal pulses as fresh carbon is dredged to the surface.  Consequently, the C/O ratio changes as the star moves up the AGB, from C/O$\sim$0.5 in M stars to C/O$\sim$1.0 in C stars.  Figure \ref{syncofield} shows examples of the C$_2$ region (at 5140\AA) and the O I forbidden line (at 6300\AA), used to derive the C and O abundances.  Syntheses with no C or O present are shown as the short dashed lines.  The other synthetic spectra show [C/Fe] at varying enhancements and [O/Fe]=0.00, +0.10, +0.25 and +0.40.

\begin{figure}
\plotone{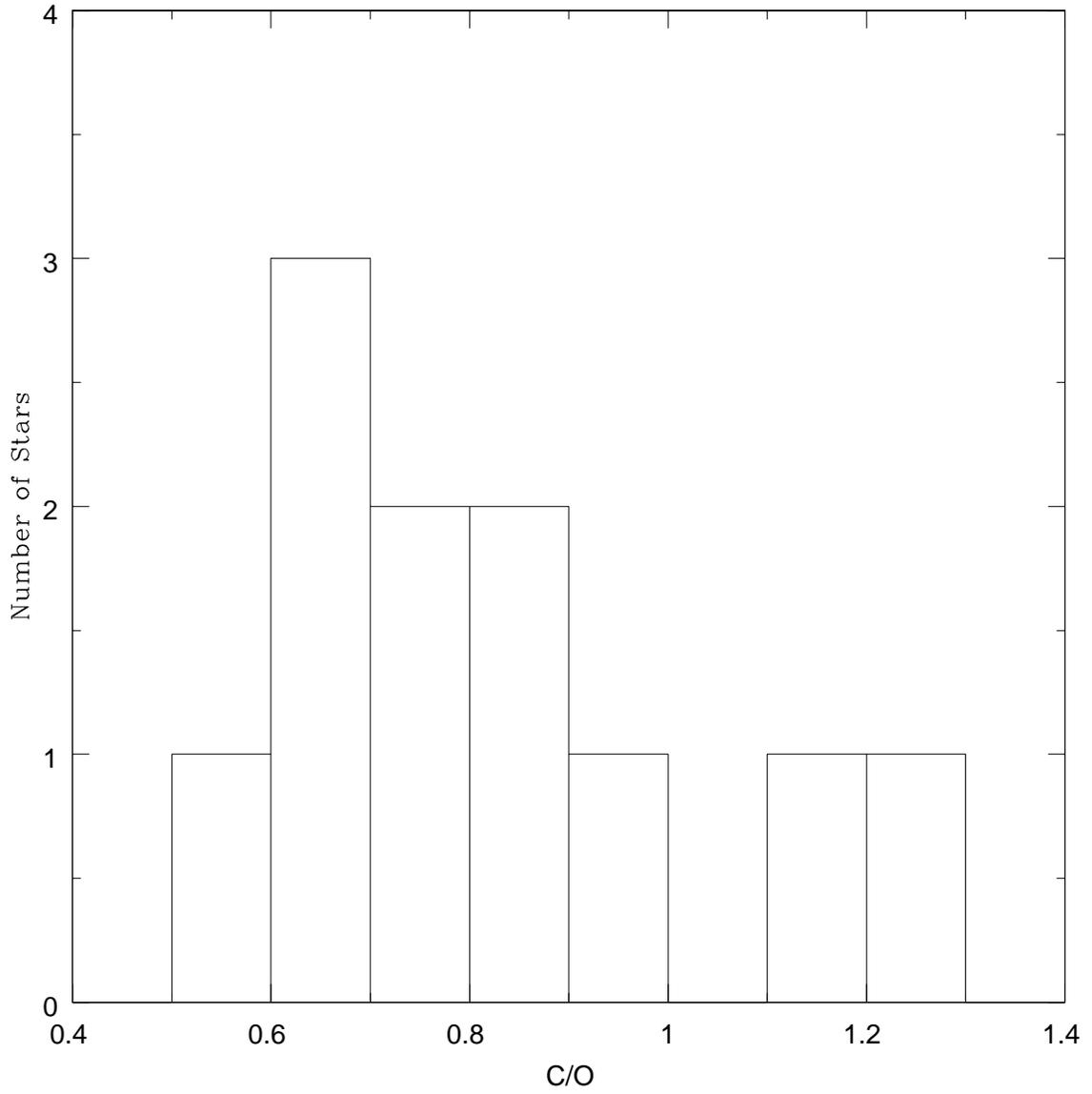}
\caption{Distribution of C/O ratio for the field stars studied.} \label{cohist}
\end{figure}

As can be seen in Tables \ref{mstars} and \ref{sstars}, most stars show enhancements in C from [C/Fe]=+0.11 to +0.79 depending on spectral type, as expected.  Oxygen was also observed to be either near-solar or slightly enhanced, from [O/Fe]=0.00 to +0.30, again depending on spectral type.  Generally speaking, the two elements are enhanced by similar amounts in any one star, with stars greatly enhanced in C also showing slight enhancements in O.  The C/O ratios are displayed in Figure \ref{cohist}, with C/O shown as a function of the frequency distribution among the sample.  The C/O ratio is an indication of the quantity of carbon brought to the surface by the third dredge-up process.  As can be seen from Table \ref{mstars}, the M stars have C/O ratios ranging from C/O=+0.47 to +0.79 with a mean of C/O=+0.59.  Table \ref{sstars} show the S and SC stars values ranging from C/O=+0.83 to +1.32 with a mean of C/O=+1.00.  There is quite clearly a distinct spread in the C/O values in this sample. 
 
 \subsection{s-process element abundances}

The M stars show small or no enhancements of s-process elements, Y, Zr, La and Nd.  Four M stars give mean values of [ls/Fe]=+0.08 and [hs/Fe]=+0.09, while the other outlier, HD7351(M3) gives more enhanced values of [ls/Fe]=+0.45 and [hs/Fe]=+0.46.  These four M stars can be said to show minimal s-process element enhancements.  This is consistent with theory, in that there is not expected to be any process in M stars that would cause enhanced s-process element abundances as there have not been any thermal pulses this early in the AGB phase.  The star HD7351 also displays a higher C/O ratio, which suggests it is a more evolved spectral class, discussed further in later sections.  Of the eight stars in common with previous studies (Smith and Lambert, 1985; Smith and Lambert, 1986; Abia and Wallerstein, 1998), most s-process element abundances agree to within 0.18 dex.  The greater the observed enhancement the more the divergence in agreement between studies, although abundances in this study never disagree with previous studies by more than 0.25 dex.  There also appears to be no systematic offsets in this analysis as abundances in this study are equally distributed both higher and lower than previous published values.  

\begin{figure}
\plotone{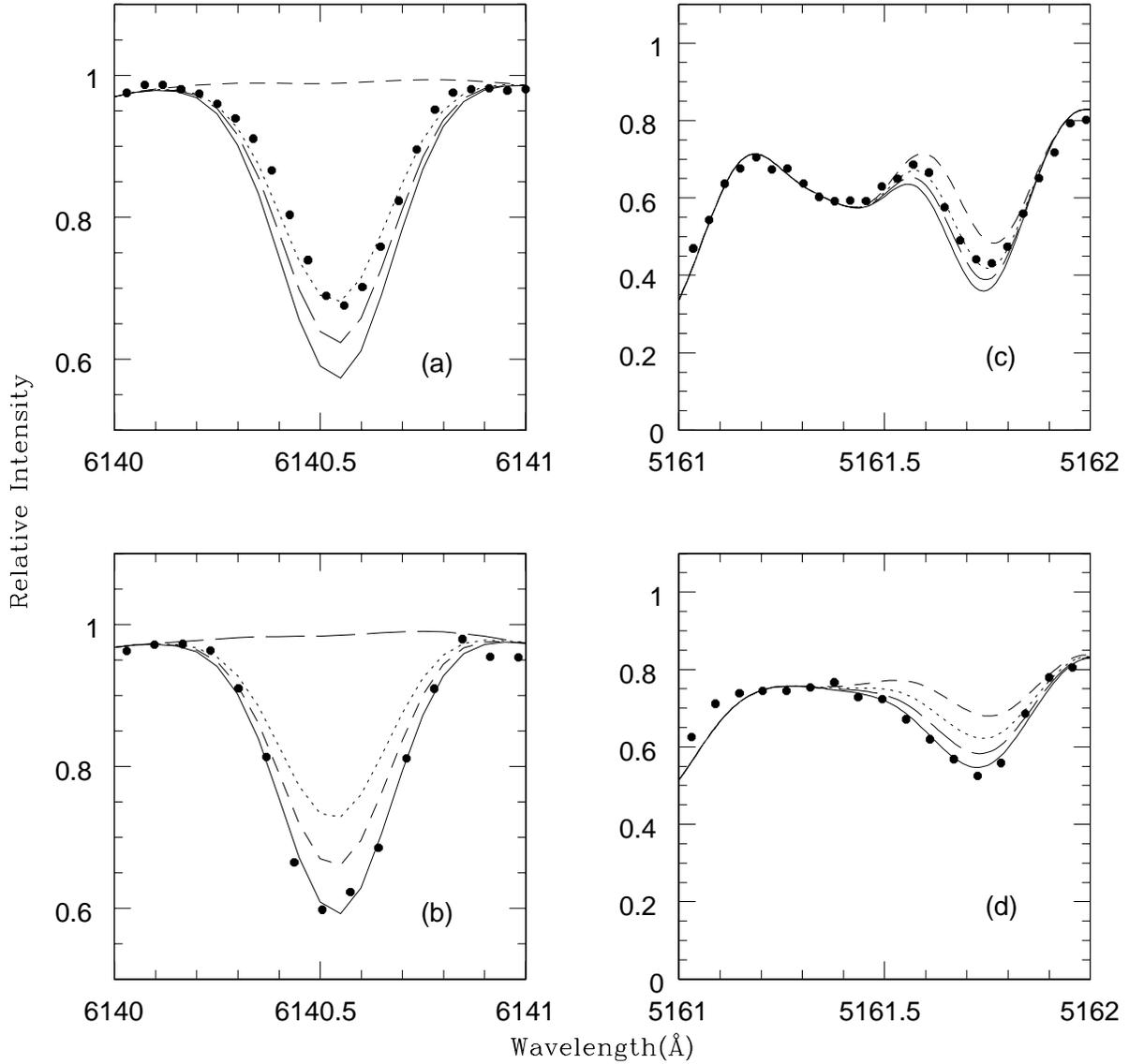}
\caption{Examples of line synthesis used to ascertain Zr and Nd abundances.  a) HD49331 and b) HD64332 with Zr abundances of no Zr (short dashed), [Zr/Fe]=0.00 (dotted), 0.50 (long dashed) and 1.00 (solid); c) HD49331 and d) HD64332 with Nd abundances of no Nd, [Nd/Fe]=0.00, 0.50 and 1.00.}\label{synzrfield}
\end{figure}

\begin{figure}
\plotone{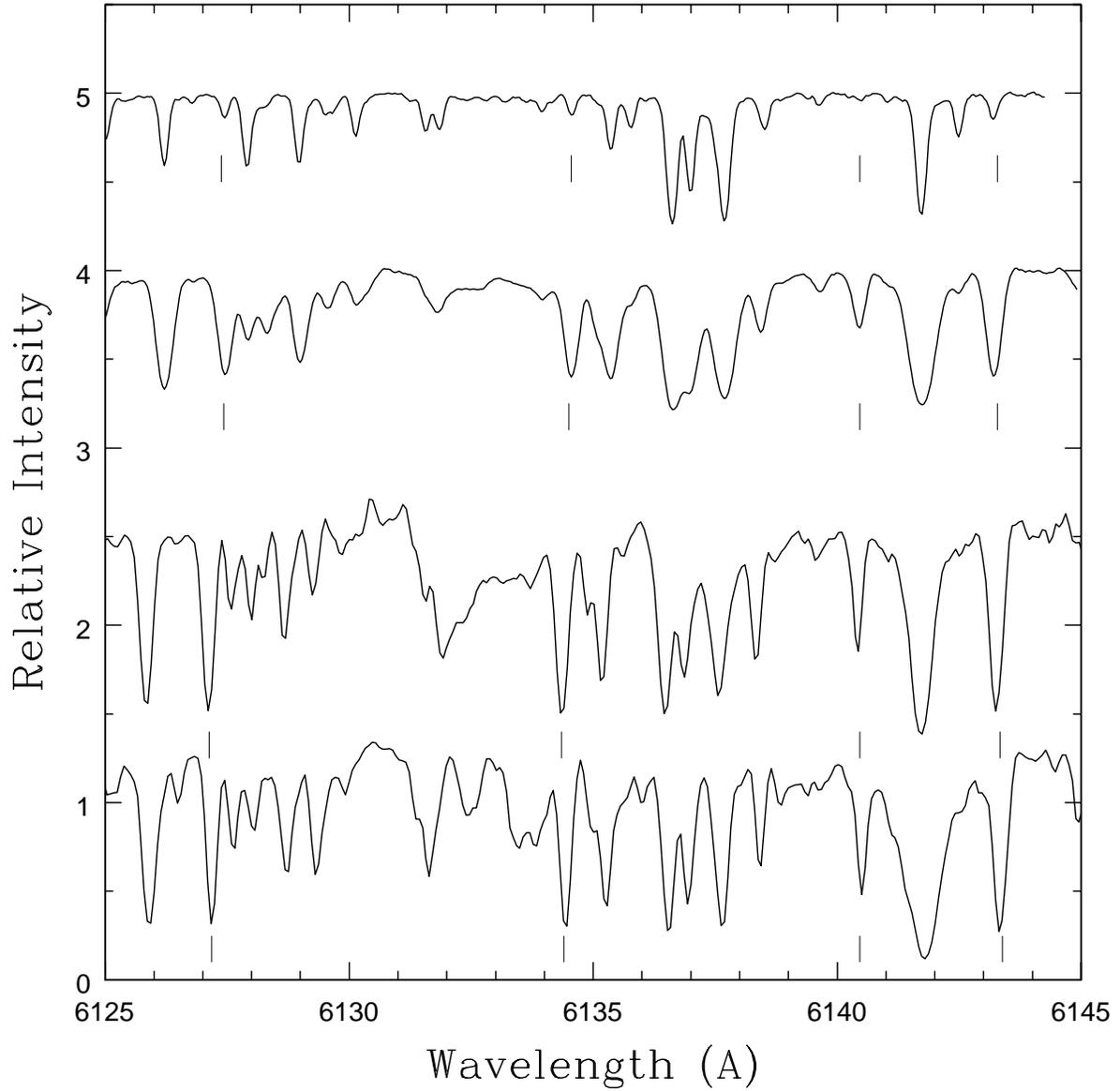}
\caption{Spectra of a K giant (top) (HD27371, T$_{eff}$=5200K, log g=3.0, [Fe/H]=0.10), and an M (HD49331), S (HD49368) and C (HD30959) star.  The increase in the strength of the Zr lines is obvious.} \label{stackzrfield}
\end{figure}

\begin{figure}
\plotone{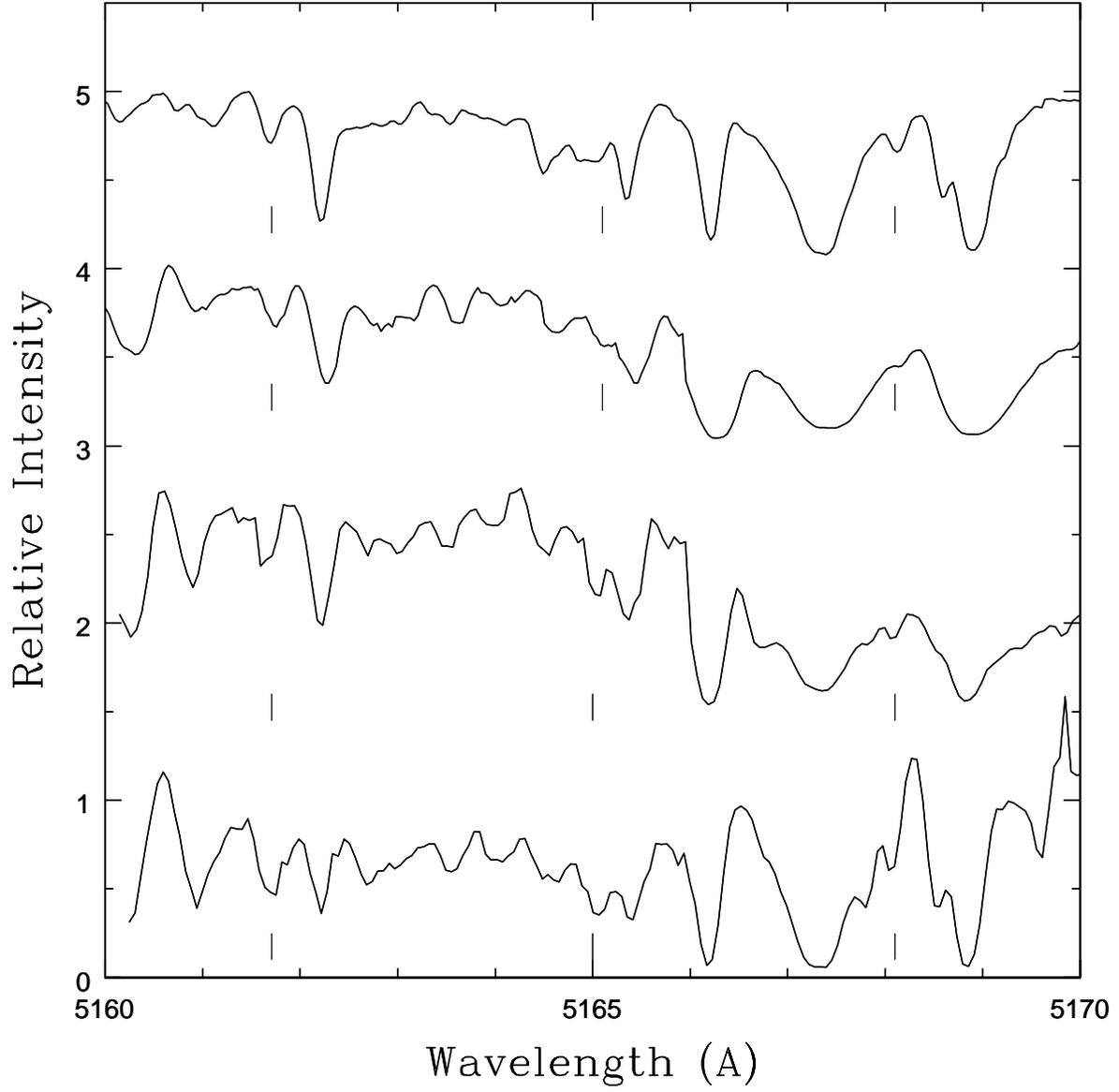}
\caption{Spectra of the same stars shown in Figure \ref{stackzrfield} (top to bottom: HD27371, HD49331, HD49368 and HD30959).  The increase in the strength of the Nd lines is obvious.} \label{stackndfield}
\end{figure}

Figure \ref{synzrfield} shows examples of a Zr I (at 6140\AA) and Nd II line (at 5161\AA), used to derive the Zr and Nd abundances.  Syntheses with no Zr or Nd present are shown as the short dashed lines.  The other synthetic spectra show [X/Fe]=0.00 and enhancements of +0.5 and +1.0 dex.  It is clear that the M star (3a, 3b) shows no Zr or Nd enhancements while the C star (3c, 3d) shows increasingly strong Zr and Nd lines.  Figures \ref{stackzrfield} and \ref{stackndfield} show schematics of slightly larger regions and can be used to compare the relative strengths of Zr and Nd lines in a K giant (top: HD 27371) compared to M (HD 49331), S (HD 49368) and C (bottom: HD 30959)) stars.

There is a definite separation, marked by an increase in the s-process elements in the S stars.  The S and C stars show mean values of [ls/Fe]=+0.76 and [hs/Fe]=+0.65.  A single star, HD216672 (S5) has considerably lower values of [ls/Fe]=+0.39 and [hs/Fe]=+0.39 dex.  With its lower s-process element enhancements and lower C/O ratio, it and HD7351 lie between the two extreme groups of M and C stars.

\subsection{Uncertainties on derived abundances}

All abundances quoted in this paper carry uncertainties, some of which are more significant than others.  There are three main sources of uncertainty in these derived abundances, discussed here in detail. 

The main source of uncertainty is the spread in derived abundances from multiple lines of the same element.  While, in theory, this should be small, different lines of the same element can give different abundances, by as much as $\sim$0.2 dex.  This can be due to incorrect log gf values, although in this study a reverse analysis using Arcturus was done for all lines used to avoid precisely this problem (see Section \ref{obsan} for details).  Non-LTE effects, blending and data quality may also contribute to this problem.  It is this value that is quoted as the $\sigma$ value in Tables \ref{mstars} and \ref{sstars}.  Where abundances are obtained from only one line (C, O, Eu and occasionally s-process elements) the $\sigma$ value is the uncertainty from a single line measurement (discussed below).

The second source of uncertainty is the accuracy with which it is possible to determine an abundance from any particular individual line.  This uncertainty can vary markedly depending on the line strength, any blending or crowding in the region and the overall quality of the data.  In this study it is estimated that this value should be no more than $\sim0.15$dex.

The final source of uncertainty is the difference in abundances obtained using different atmospheric parameters in the stellar models.  This depends solely on how well the atmospheric parameters can be constrained.  Table \ref{abdep} shows the dependence of derived abundances on the chosen atmospheric parameters.  These values were obtained by choosing equivalent width values which gave abundances representative of those obtained via spectrum synthesis and then running an abundance analysis on these equivalent widths with four different models, with $\Delta$T$_{eff}$=+250K, $\Delta$log g=+0.3 and $\Delta$$\xi$=+0.25km s$^{-1}$.  These $\Delta$ values represent higher uncertainties on the atmospheric parameters than the ones estimated in this study (see Table \ref{fsap}).  It can be seen that all atmospheric parameters influence different lines, stressing the importance of ascertaining a correct model atmosphere at the outset. 

Two elements which merit specific discussion are C and O.  Given the importance of these two elements to AGB modeling, it is necessary to provide a realistic estimate of the uncertainty involved in these abundance derivations.  The abundances for C and O are derived from one region or line only so the uncertainty associated with these abundances depends on how well that particular feature can be measured, as opposed to an abundance spread over various features.  It is estimated that the C and O abundances can be derived to within $\pm$0.15 dex on these particular features (examples shown in Figure \ref{syncofield}).  This has a dramatic effect in the uncertainty in the quoted C/O ratios.  Taking into account the 0.15 dex uncertainty in both C and O, it results that the uncertainty on the final C/O ratio is $\pm$0.30 dex.  While this is a large quoted uncertainty it is hoped that, in reality, the C/O value has an uncertainty less than this.  Data of higher quality and measurement of more features of C and O, including molecular features out of wavelength range of this investigation, would refine this value further.  In addition, these values are absolute uncertainties and the relative uncertainties between the stars should be smaller due to the consistent data sets and analysis.

\section{Discussion}\label{discuss}

\subsection{General trends in AGB stars}

\begin{figure}
\plotone{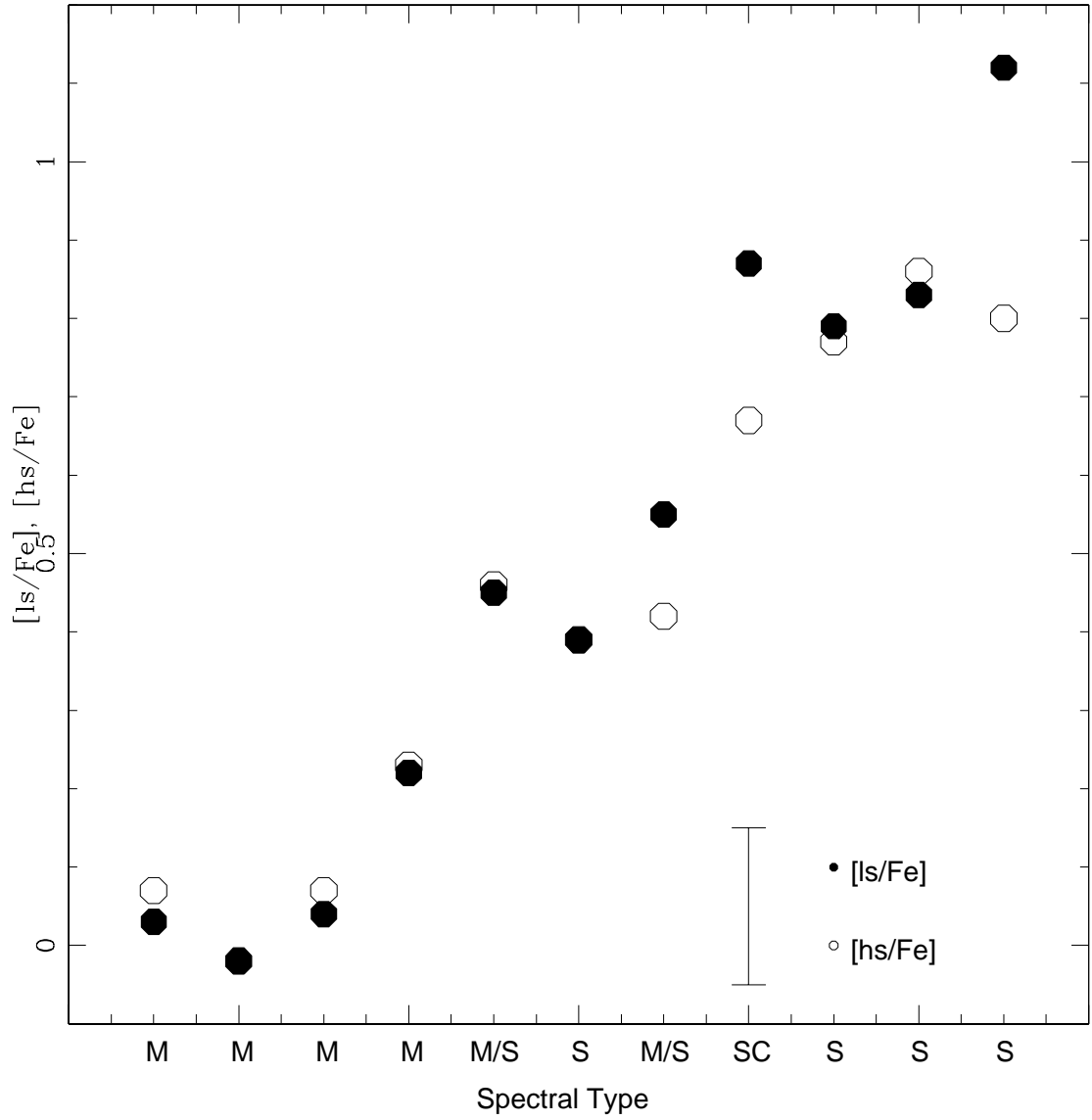}
\caption{Trends in [ls/Fe], [hs/Fe] and [hs/ls] for different spectral types.} \label{starhsls}
\end{figure}

\begin{figure}
\plotone{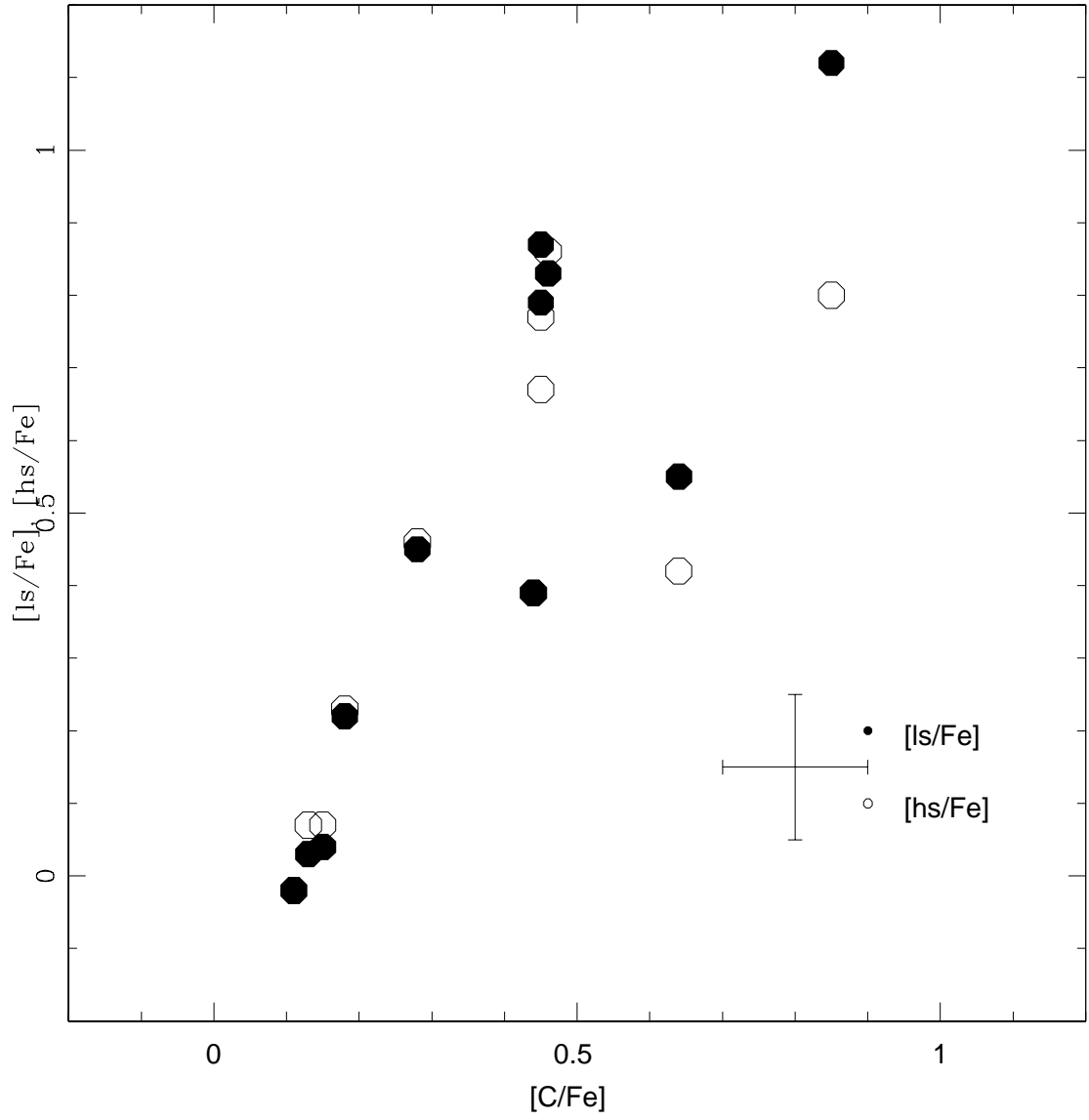}
\caption{Trends in [ls/Fe], [hs/Fe] as the [C/Fe] value increases.} \label{chsls}
\end{figure}

\begin{figure}
\plotone{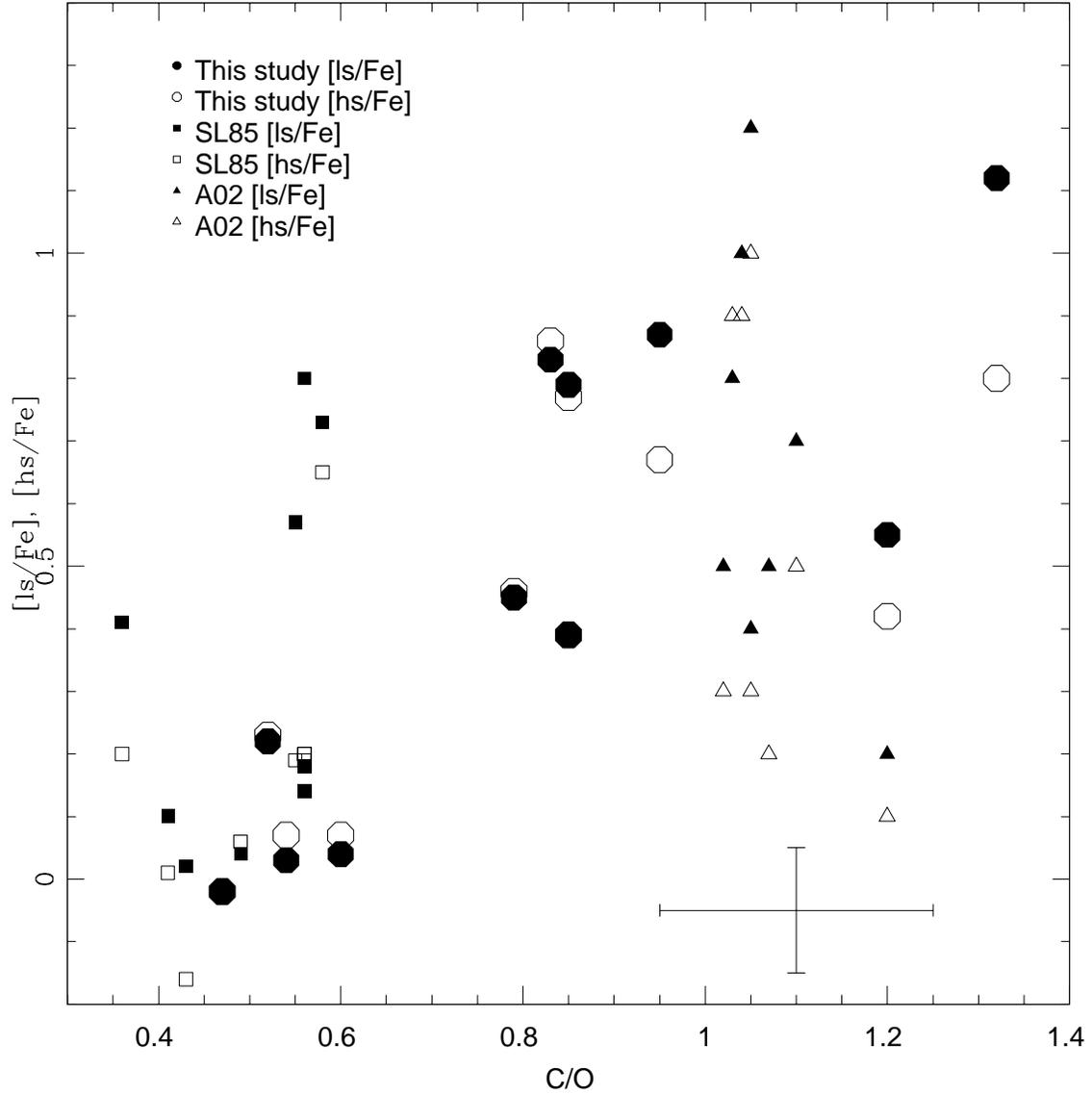}
\caption{Trends in [ls/Fe], [hs/Fe] and [hs/ls] as the C/O ratio increases.  Results from previous studies are also shown (squares: Smith and Lambert, 1985; triangles: Abia et al., 2002).} \label{cohsls}
\end{figure}

Generally, it is expected that as the star moves up the AGB the [C/Fe] value and C/O ratio increase due to the C$^{12}$ brought to the surface by the third dredge-up process.  Another important by-product of this phenomenon is the expected surface enhancements of the s-process elements.  In an attempt to quantify this relationship, this research examines the [ls/Fe] and [hs/Fe] values as a function of increasing AGB spectral type (Figure \ref{starhsls}), [C/Fe] value (Figure \ref{chsls}) and C/O ratio (Figure \ref{cohsls}). 

Figure \ref{starhsls} shows the enhancements of the light s- (as filled circles) and heavy s- process elements (as open circles) for each individual star as the AGB phase progresses.  The M stars are clearly minimally enhanced, and as the spectral type advances the enhancements of the light and heavy s-process elements clearly increases.   However, this figure relies heavily upon spectral classification, which can often be difficult or ambiguous.  

A more logical way of quantifying s-process enhancements is through the [C/Fe] and C/O values.  Figure \ref{chsls} shows the [ls/Fe] (as filled circles) and [hs/Fe] (as open circles) values as a function of the derived [C/Fe] value.  In general, it is clear that as the [C/Fe] value increases s-process enhancements are seen.  All stars seem to follow this trend between [C/Fe] and [s/Fe].  However, there is some indication that the s-process elements may be enhanced first, followed by the carbon enhancement.  This is most clearly seen in Figure \ref{chsls}, in which there is a steep rise in [ls/Fe] and [hs/Fe] at [C/Fe] = 0.10 to 0.50 and then a plateau region from [C/Fe] = 0.50 to 1.0, at which the s-process elements have an approximately constant enhancement of $\sim$1.0.  Although this generalization is based is on a relatively small number of objects, it may provide some observational test data for the modelers of the TDU.  It should be noted that this sample most likely includes a mix of stellar masses and this is a critical parameter for the efficiency of the TDU episodes and, consequently, the C and O abundances.  It is very difficult to obtain accurate masses for field stars, and thus no estimate of stellar mass has been attempted here.  This problem can be avoided by analysing AGB stars in a globular cluster, where masses are more accurately determined although few studies have been published (Wylie et al., 2006; Yong et al., 2008).

Figure \ref{cohsls} shows the [ls/Fe] (as filled circles) and [hs/Fe] (as open circles) values as a function of C/O ratio.  This ratio is also an indication of the amount of dredge-up occurring in the star and is expected to increase from $\sim$0.5 at the beginning of the AGB phase to $\sim$1.0 at the end of the thermally pulsing AGB stage.  Results from several stars in previous studies (Smith and Lambert, 1985; Abia et al. 2002) that are not in common with this paper are also included as squares and triangles respectively.  

As can be seen in Figure \ref{cohsls}, generally all stars follow the expected trend of increasing [s/Fe] with increasing C/O and the spread does cover the expected range of C/O=0.5 for M stars to C/O$\geq$1.0 for C stars.  The plateau effect at [ls/Fe] and [hs/Fe] = +1.0 is also evident.  Figure \ref{cohsls} also shows how well this current study agrees with trends seen in previous studies.  The stars from Smith and Lambert (1985), shown as squares, and the stars from Abia et al. (2002), shown as triangles, seems to show no marked discrepancies from the trend observed in this study.  The inclusion of these 17 stars, which results in a larger overall sample, strongly supports the general conclusions drawn from the smaller sample studied in this paper.  

Another result that Figure \ref{cohsls} suggests is that as the C/O ratio increases the difference between the [ls/Fe] and [hs/Fe] enhancements gets larger.  This can be confirmed by looking at the individual [hs/ls] values as a function of C/O.  For the three stars with C/O $\geq$ 0.95 their average [hs/ls] value is -0.22, whereas the average [hs/ls] for all other stars is -0.02.  This may suggest that initially, while the s-process and carbon enhancements are minimal they are all enhanced by the same amounts.  However, as the enhancements increase in value, the two s-process element peaks separate and begin to be enhanced at different rates, with the light s-process peak being favoured.  For near-solar metallicity it is predicted that the light s-process elements are more enhanced than the heavy s-process elements (see Busso et al. (2001) and references therein), and earlier studies have also suggested this trend observationally (Smith and Lambert, 1985; Abia et al. 2002).  This prediction is now further confirmed by these present results.  

\subsection{r-process, Eu}

As can be seen in Tables \ref{mstars} and \ref{sstars}, three stars, HD44478 (M3), HD112300 (M3) and HD216672 (S5) show slight deficiencies in the r-process element, Eu, while three others, HD49368 (S3), HD64332 (S4) and HD286340 (SC7) show possible small enhancements in Eu (of only [Eu/Fe]$\sim$0.1dex), which are not predicted by the theoretical modelling.  These enhancements are small and within uncertainties the [Eu/Fe] ratio is compatible with solar ratios.  However, if these enhancements are genuine, and not a product of uncertainty on measurements, this raises an interesting issue.  The AGB evolutionary phase is not expected to produce r-process elements in any way, so it is not expected that AGB stars would show Eu enhancements due to their stellar evolution.  These observed enhancements may be more likely to come from the primordial enrichment of the material from which the stars formed or external pollution from other sources. 

If this Eu enhancement is genuine, it is possible that these three stars have undergone pollution from material rich in r-process elements.  This may have involved extrinsic pollution from supernova remnants or other ejecta in which r-process element formation has previously occurred.  This may explain any observed Eu enhancement in these stars.  It may also help to explain the observed Nd enhancement in these stars, which in each of these three stars shows a greater enhancement than the other heavy s-process element, La.  Nd has five stable isotopes, four of which are formed by both the s- and r-process and one of which is r-process only.  If these two stars had undergone pollution from r-process element rich material, then it is possible that the source of this pollution is responsible for the observed enhancement of Nd.  La, which has one stable isotope, can also be produced by both the s- and r-process, thus providing more support for this theory.  The large enhancements of La, Nd and the slight enhancement in Eu observed in these stars may not be due solely to AGB nucleosynthesis, but partially due to pollution from previously r-processed matter.

\section{Conclusion}

In this paper the results of an internally accurate and self-consistent abundance analysis of AGB stars are presented.  The relationship between [C/Fe], C/O and s-process element enhancements is quantified.  All general trends seen in this small sample of stars are strongly supported by those found in previous studies.  While absolute uncertainties on the crucial elements C and O are reasonably large at $\pm$0.30 dex, the relative uncertainties should be smaller due to consistent data sets and analysis methods, thus supporting the strength of the results from this internally consistent analysis.  

\acknowledgements

ECW acknowledges the support of a University of Canterbury Research Award, a Departmental Scholarship and the Dennis William Moore Scholarship, all held for the duration of this research.  This research was partially supported by a grant from the American Astronomical Society.  This research has made use of the SIMBAD database, operated at the Centre de Don\'{e}es Astronomiques de Strasbourg (CDS), France.  This publication makes use of data products from the Two Micron All Sky Survey, which is a joint project of the University of Massachusetts and the Infrared Processing and Analysis Center/California Institute of Technology, funded by the National Aeronautics and Space Administration and the National Science Foundation. 

{\it Facilities:} \facility{AAT}

\clearpage


\begin{deluxetable}{lccccccc}
\tabletypesize{\scriptsize}
\tablecaption{Basic stellar quantities.\label{starquant}}
\tablewidth{0pt}
\tablehead{
\colhead{Star} & \colhead{Spectral Type\tablenotemark{a}} & \colhead{B} & \colhead{B} & \colhead{K} &
\colhead{MBol} & \colhead{Variable Type\tablenotemark{b}} & \colhead{Tc detection\tablenotemark{c}} 
}
\startdata

HD7351   & M3 / S        & 8.07   & 6.38  & 1.64          & -3.02  & SR & no\\
HD30959  & M3 / S        & 6.53   & 4.75  & -0.66         & -3.87  & SRB (30 days) & yes\\
HD35155  & S             & 8.64   & 6.87  &  2.14         & -4.45  & EA/GS/WD & no\\
HD44478  & M3            & 4.53   & 2.91  & -1.86         & -3.31       & LB & ... \\
HD49331  & M2 / S        & 6.91   & 5.11  &  0.56         & -4.12      & susp & no \\
HD49368  & S3            & 9.53   & 7.78  & 2.46          & -3.57        & SRB & no\\
HD64332  & S4            & 9.34   & 7.64  & 2.31          & -2.42        & LB &  yes\\
HD102212 & M1            & 5.58   & 4.05  &  0.16         & -2.10     & SRB & ...\\ 
HD112300 & M3            & 4.96   & 3.38  & -1.19         & -2.37        & susp & ...\\ 
HD216672 & S5            & 8.21   & 6.47  & 1.04          & -3.42        & SRB (50 days)& yes\\ 
HD286340 & SC7           &    ...    &  ...     &   ...            &   ...           & SRB (370 days)& ...\\ 
\enddata
\tablenotetext{a}{\cite{b10}}
\tablenotetext{b}{\cite{Samus04}- SR: semi-regular, SRB: semi-regular with poorly defined period, EA/GS/WD: eclipsing binary with both giant and white dwarf component, LB: slow irregular, susp: only suspected variable, type not confirmed. }
\tablenotetext{c}{\cite{SL88}}
\end{deluxetable}

\clearpage
\begin{deluxetable}{lccclccc}
\tabletypesize{\scriptsize}
\tablecaption{Line list of all Fe I and Fe II lines used to derive abundances via equivalent width analysis.\label{felinelist}}
\tablewidth{0pt}
\tablehead{
\colhead{Species} & \colhead{Wavelength} & \colhead{$\chi$} & \colhead{log gf} & \colhead{Species} &
\colhead{Wavelength} & \colhead{$\chi$} & \colhead{log gf} \\
\colhead {} & \colhead{(\AA)} & \colhead (eV) & \colhead {} & \colhead {} & \colhead {(\AA)} & \colhead {(eV)}
}
\startdata
Fe I & 5044.21 &      2.85&    -2.150   &  &5859.56 & 4.55 & -1.080\\                     
& 5068.77  &     2.94 &   -1.230     &  &5862.33    &   4.55   & -0.600            \\           
 & 5074.75   &    4.22   & -0.200     &&5929.68      & 4.55    &-1.410                         \\
  &5159.06     &  4.28&    -0.820      &&6065.48&       2.61 &   -1.530                        \\
  &5215.18&       3.27  &  -0.870      &&6094.12  &     4.65   & -1.640                         \\
  &5618.63  &     4.21   & -1.260   &&6096.66    &   3.98    &-1.830                         \\
  &5624.02    &   4.39&    -1.330       &  &6107.91      & 4.14&    -0.191                        \\
  &5633.95 &      4.99  &  -0.120     &&6226.74&       3.88  &  -2.220                         \\
  &5635.82   &    4.26   & -1.740       &&6240.64  &     2.22   & -3.388                        \\
  &5662.50      & 4.12&    -0.570        &&6380.74    &   4.19    &-1.401                         \\
  &5686.51&       4.55  &  -0.450     &  Fe II &   5197.56&       3.23&    -2.100                \\        
  &5691.50  &     4.30   & -1.370    &&5234.62&       3.22&    -2.050                        \\
  &5705.47    &   4.30&    -1.360         &      &5534.85  &     3.24  &  -2.740                         \\
  &5717.83      & 4.28  &  -0.980          &&6084.10    &   3.20   & -3.908                         \\
  &5731.76&       4.26   & -1.150   &      &6238.39      & 3.89&    -2.630                         \\
  &5806.73  &     4.61&    -0.900      &&6247.55 &      3.89  &  -2.516                         \\
                       &5809.22    &   3.88  &-1.690 &     &6432.68   &    2.89   & -3.808                         \\
                    &5852.22      & 4.55   & -1.180     &      &6456.39     &  3.90    &-2.275  \\                      
  \enddata
  \end{deluxetable}

\clearpage
\begin{deluxetable}{lccclccc}
\tabletypesize{\scriptsize}
\tablecaption{Line list of all other lines used to derive abundances.\label{otherlinelist}}
\tablewidth{0pt}
\tablehead{
\colhead{Species} & \colhead{Wavelength} & \colhead{$\chi$} & \colhead{log gf} & \colhead{Species} &
\colhead{Wavelength} & \colhead{$\chi$} & \colhead{log gf} \\
\colhead {} & \colhead{(\AA)} & \colhead {(eV)} & \colhead {} & \colhead {} & \colhead {(\AA)} & \colhead {(eV)}
}
\startdata
O I    & 6300.30 & 0.00 & -9.75 &     &  6140.46 & 0.52 & -1.41    \\
      &  6363.78 & 0.02 & -10.25    &  & 6143.18 & 0.07 & -1.10\\
  Y I  &   5526.72 & 2.00 & -0.65    &  Zr II& 5112.28 & 1.66 & -0.59\\
     & 6435.05 & 0.07 & -1.02& &  5350.09 & 1.83 & -0.94 \\
Y II        & 5087.43 & 1.08 & -0.17& & 5350.36 & 1.77 & -1.18   \\
& 5119.12 & 0.99 & -1.36 &     La II & 5114.56 & 0.24 & -1.06  \\          
      & 5473.39 & 1.74 & -1.02           & & 5122.99 & 0.32 & -0.93  \\  
            & 5544.62 & 1.90 & -1.09     &  &  5797.60 & 0.24 & -1.51     \\
& 5546.03 & 1.75 & -1.10      && 5805.77 & 0.13 & -1.61  \\
Zr I        & 4805.89 & 0.69 & -0.57 & &5808.31 & 0.00 & -1.56    \\
& 4815.06 & 0.65 & -0.53 &Nd II &5161.71 & 0.74 & -0.98 \\          
      & 4815.64 & 0.60 & -0.13      && 5165.13 & 0.68 & -0.06 \\ 
      & 4828.06 & 0.62 & -0.64    &  &5167.92 & 0.56 & -0.98 \\    
 & 4809.48 & 1.58 & +0.16 & & 5795.15 & 1.26 & -1.13 \\
       & 6127.48 & 0.15 & -1.06       & &    5804.00 & 0.74 & -0.53     \\
& 6134.57 & 0.00 & -1.28 &Eu II     &      6645.13 & 1.38 & +0.40  \\            
\enddata    
\end{deluxetable}

\clearpage
\begin{deluxetable}{lcccc}
\tabletypesize{\scriptsize}
\tablecaption{Atmospheric Parameters for Field Stars\label{fsap}}
\tablewidth{0pt}
\tablehead{
\colhead{Star} & \colhead{Ph. T$_{eff}$} & \colhead{Sp. T$_{eff}$} & \colhead{log g} & \colhead{$\xi$}\\ 
\colhead{} & \colhead{(K)} & \colhead{($\pm$ 200K)} &
\colhead{$\pm$ 0.2} &  \colhead{($\pm$ 0.25 km s$^{-1}$)} 
}
\startdata
HD7351     		 & 3740          & 3650          & 0.5   & 1.50 \\
HD30959    		 & 3650          & 3600          & 1.0   & 1.40 \\
HD35155    		 & 3670          & 3700          & 1.0   & 1.30 \\
HD44478    		 & 3850          & 3900          & 1.0   & 1.80 \\
HD49331    		 & 3640          & 3500          & 0.5   & 1.25 \\
HD49368    		 & 3700          & 3700          & 1.0   & 2.00 \\
HD64332    		 & 3750          & 3600          & 1.0   & 1.50 \\
HD102212   		 & 3970          & 4000          & 1.5   & 1.40 \\ 
HD112300   		 & 3890          & 3900          & 0.5   & 1.10 \\ 
HD216672   		 & 3700          & 3700          & 1.0   & 1.20 \\ 
HD286340   		 &   ...             & 3500          & 1.0   & 1.50 \\
\enddata
\end{deluxetable}

\clearpage

\begin{deluxetable}{lcccccccccc}
\tabletypesize{\scriptsize}
\tablecaption{Element Abundances for field M and MS stars\label{mstars}}
\tablewidth{0pt}
\tablehead{
\colhead{X/H} & \colhead{HD7351} & \colhead{$\sigma$} & \colhead{HD44478} & \colhead{$\sigma$} &
\colhead{HD49331} & \colhead{$\sigma$} & \colhead{HD102212} &
\colhead{$\sigma$} &  \colhead{HD112300} & \colhead{$\sigma$} 
}
\startdata
Fe I   & 0.05    &     0.08      &  0.00    &0.09           &0.15    &  0.29         &  -0.04   &  0.12        &  0.02    &0.05          \\
Fe II  & 0.02    &     0.10      &  0.00    &  0.08         &0.22    &	 0.06        &  -0.04   & 0.08         &  0.00    &  0.06        \\	    \hline
[X/Fe] & & & & & & & & & &\\ 
C      & 0.28    &     0.15      &  0.15    &0.15          &  0.13  &0.15          & 0.18     &0.15         & 0.11     &0.15       \\ 
O      & 0.01    &       0.15    &  0.00    &   0.15        & 0.03   &   0.15       & 0.08     &   0.15      & 0.07     &   0.15      \\
Y I    & 0.29    &     0.15    &  0.04    &    0.03       & 0.16   & 	0.03         & 0.14     & 0.15        &  -0.01   & 0.15         \\  
Y II   & 0.39    &   0.13        &  0.06    &  0.02         & 0.10   &	 0.03        & 0.16     &  0.10        &  -0.01   &   0.10    	 \\ 
Zr I   &  0.50   &    0.24       &  0.06    &  0.03         & 0.16  &     0.06      &   0.26   &  0.09        &  -0.08   &     0.02     \\
Zr II  &  0.51   &   0.21        &  0.03    &  0.07         &0.11   &     0.07      &   0.29   &   0.10       &   -0.01   &         0.07 \\
La II  &  0.43   &   0.07        &  0.05    &   0.04        &0.08   &0.01	         &   0.27   &  0.04        &  -0.02   & 0.15         \\     
Nd II  &  0.49   &   0.07        &  0.09    &    0.03       &0.07   &   0.15       &   0.19   &   0.07       &   -0.01   &         0.09 \\
Eu II  &  0.05   &  0.10         &  -0.11    &0.10           &-0.01    &	0.10         &   0.02   &0.10          &  -0.03   &      0.10  \\ \hline
C/O    & 0.79    &    ...       & 0.60     &  ...         & 0.54   &  ...         & 0.54     &...          & 0.47     &...          \\
ls     &  0.45   &   ...        &  0.04    &        ...   &0.08   &	      ...   & 0.22	    &  ...        &   -0.02   &  ...        \\    
hs     &  0.46   &     ...      &  0.07    &          ... &0.07  &	         ...& 0.23	    &	  ...     &   -0.02   &     ...     \\    
hs/ls  &  0.01  & ...          &  0.02    &           ...& -0.01   &           ...&  0.01   &        ...  &  0.00    &        ...\\ 
\enddata
\end{deluxetable}

\clearpage
\begin{deluxetable}{lcccccccccccc}
\tabletypesize{\scriptsize}
\tablecaption{Element Abundances for field S and SC stars\label{sstars}}
\tablewidth{0pt}
\tablehead{
\colhead{X/H} & \colhead{HD30959} & \colhead{$\sigma$} & \colhead{HD35155} & \colhead{$\sigma$} &
\colhead{HD49368} & \colhead{$\sigma$} & \colhead{HD64332} &
\colhead{$\sigma$} &  \colhead{HD216672} & \colhead{$\sigma$} & \colhead{HD286340} & \colhead{$\sigma$}
}
\startdata
Fe I   &  0.02   &  0.22         &   -0.12  & 0.11         &  -0.20   &  0.10        &-0.25      & 0.18         & 0.01     &  0.11        & 0.00     &  0.13         \\
Fe II  & 0.07   &	0.16     &   -0.10  & 0.10         &  -0.20   & 0.13         &-0.27      &   0.21     & 0.06     &    0.10      & 0.00     &   0.12       \\      \hline
 $[$X/Fe$]$& & & & & & & & & \\ 
C      & 0.58    & 0.15          &   0.45   & 0.15        &   0.46   &  0.15        & 0.79            & 0.15  & 0.44     &  0.15       & 0.45     &         0.15 \\
O      & 0.13    & 0.15          &   0.15   &    0.15      &   0.17   &    0.15      & 0.30   &      0.15 & 0.14     &   0.15      & 0.10     &    0.15\\
Y I    &  0.78   & 0.20	     &	 0.77   &     0.07     &   0.85   &   0.07       & 1.05   &    0.14            & 0.49     &   0.40       & 0.80     &    0.10       \\ 
Y II   &  0.66   &	     0.06&	 0.83   &    0.05      &   0.82   &       	0.09 & 1.12  &         0.09      & 0.34     &        0.14  & 0.76     &    0.20      \\       
Zr I   &   0.56  &         0.06  &   0.84   &   0.10       &   0.82   & 0.08         & 1.20   &         0.09     & 0.53     &        0.11  & 0.98     &       0.17  \\
Zr II  &  0.43  & 0.10          &   0.75  &     0.07     &   0.85   &     0.07     & 1.12  &  0.07            & 0.44     &  0.12        & 0.95     &   0.07    \\
La II  &  0.43   &	  0.20   &   0.77   &    0.06      &   0.84   &    0.05      & 0.75   &     0.09          & 0.24     &  0.15        & 0.62     &     0.09  \\         
Nd II  &  0.40   &       0.10    &   0.78   &    0.05      &   0.87   &     0.05     & 0.85       &  0.05      & 0.54     &   0.06       & 0.71     & 0.08       \\
Eu II  & 0.02   &0.10	     &   0.00   &0.10          &   0.09   & 0.10         & 0.06   &0.10                 & -0.07    &0.10          & 0.10     &0.10      \\ \hline
C/O    & 1.20    & ...          &   0.85   &  ...        &   0.83   &   ...       & 1.32   &  ...             & 0.85     &    ...      & 0.95     &    ...   \\
ls     &  0.55   &	     ...&	 0.79	&         ... &   0.83   &          ...& 1.12   &         ...          & 0.39     &       ...   & 0.87     &       ... \\        
hs     &   0.42  &	 ...    &	 0.77	&	   ...&   0.86   &          ...& 0.80   &            ...        & 0.39     &         ... & 0.67     &         ... \\      
hs/ls  &  -0.13  &   ...        &  -0.02   &         ... &  0.02    &         ... & -0.32  &          ...  & 0.00     &          ...& -0.21    &   ... \\ 
\enddata
\end{deluxetable}

\clearpage
\begin{deluxetable}{lccc}
\tabletypesize{\scriptsize}
\tablecaption{Dependence of derived abundances on atmospheric parameters\label{abdep}}
\tablewidth{0pt}
\tablehead{
\colhead{Species} & \colhead{$\Delta$ T$_{eff}$} & \colhead{$\Delta$ log g} & \colhead{$\Delta$ $\xi$} \\
\colhead{} & \colhead{(+250K)} & \colhead{(+0.3)}  & \colhead{(+0.25 km s$^{-1}$)}
}
\startdata
    O I &  -0.60&0.35  &-0.02 \\
    Y I & 0.31 & 0.02 & -0.16 \\
    Y II & -0.09 & 0.15 &  -0.16 \\
    Zr I & 0.32 & 0.03 & -0.31 \\
    Zr II & -0.13 & 0.15 & -0.11 \\
    La II & 0.02 & 0.15 & -0.24 \\
    Nd II & 0.01 & 0.14 & -0.18 \\
    Eu II & -0.05 & 0.15 & -0.07 \\ 
    \enddata
\end{deluxetable}

\clearpage


\clearpage

\clearpage

\clearpage

\clearpage

\clearpage

\clearpage

\clearpage

\clearpage

\clearpage

\clearpage

\end{document}